\documentstyle[twocolumn,pra,aps,epsfig]{revtex}
\begin{document}
\draft
\title{Nonlinear manipulation and control of matter waves}
\author{E. V. Goldstein, M. G. Moore and P. Meystre}
\address{Optical Sciences Center and Department of Physics,
University of Arizona, Tucson, AZ 85721}
\maketitle

\begin{abstract}

This paper reviews some of our recent results in nonlinear atom optics.
In addition to nonlinear wave-mixing between matter waves,
we also discuss the dynamical interplay between optical and matter waves.
This new paradigm, which is now within experimental reach,
has the potential to impact a number of fields of physics, including the
manipulation and applications of atomic coherence, and the preparation of
quantum entanglement between microscopic and macroscopic systems. Possible
applications include quantum information processing, matter-wave holography,
and nanofabrication.
\end{abstract}

\section{Introduction}
The last few years have been exciting ones indeed for atomic, molecular and
optical physics. Advances have been spectacular, both on the theoretical
and the experimental front, and many of the ideas which were ``far-off''
proposals just a few years back have now become reality. This is in
particular the case for ``atom lasers'' and nonlinear atom optics. Since the first
demonstration of a primitive atom laser by Ketterle and coworkers
\cite{MewAndKur97}, at least three other systems have been demonstrated
\cite{AndKas98,HagDehKoz99,BloHanTil99}, including a quasi-cw system
by the group of H\"ansch \cite{BloHanTil99},
and a ``mode-locked'' system by Kasevich \cite{AndKas98}. Just as important
was the experimental demonstration of matter-wave four-wave mixing by Phillips'
group at NIST \cite{denHagWen99}, and shortly thereafter of matter-wave
superradiance by Ketterle's group \cite{InoChiSta99}.
In addition, there is an exciting convergence of interests taking
place between Bose-Einstein condensation and the study of
quantum-coherent effects such as echoes \cite{KukChe98}
and electromagnetic induced transparency(EIT), demonstrated spectacularly
in the recent experiments of Hau et al \cite{HauHarDut99} and others
\cite{KasSauZib99}.

Any potential application of quantum degenerate gases will
rely on one's ability to manipulate and control them.
One obvious way to do that is by optical methods: It has long been known
that optical fields can be used as atom-optical elements \cite{AdaSigMly94},
for instance as diffraction gratings. They can work just as well for
condensates, and indeed, Phillips and coworkers \cite{KozDenHag99,denHagWen99}
used them in their trailblazing nonlinear atom optics
experiments to prepare three of the matter waves that were then used to
generate a fourth one. Optical waves can also be used to produce optical
dipole traps, which have played a central role in the studies of
multicomponent condensates by the MIT group \cite{InoChiSta99}.
As a final example, we can mention the use
of an optical lattice to build a ``mode-locked'' atom laser
by Kasevich and coworkers \cite{AndKas98}.

In this first generation of applications, the light field played
a passive role: it influenced the dynamics of the matter waves, but the
back-action of the atomic field on the optical field was largely
neglected. In recent work, we \cite{MooMey98,MooMey99,MooZobMey99}
and others \cite{ZenLinZha95,LawBig98,Kua98} have introduced a new
paradigm where both the optical and the matter waves are dynamically coupled,
independent entities.

The remainder of this paper reviews some recent results obtained by our
group in nonlinear atom optics. Section II discusses a situation where
atoms in a multicomponent condensate interact via spin-changing collisions.
We show how the dynamics of this system is characterized by the occurence
of collapses and revivals in the population of the various magnetic sublevels
involved. Section III considers the nonlinear mixing of optical
and matter waves,
and illustrates the optical control of the coherence properties of a
quantum-degenerate atomic gas. Potential applications of atom optics
are numerous, and just starting to be explored, as an outlook into
the future, Section IV briefly reviews just one such potential application,
atom holography.

\section{Nonlinear atom optics}

Several years ago, we proposed the idea of nonlinear atom optics
\cite{LenMeyWri93,LenMeyWri94}. A number of theoretical investigations
along these lines have now been carried out, including the study of
matter-wave solitons
\cite{LenMeyWri94,MorBalBur97,ReiCla97,DumCirLew98,JacKavPet98,ZobWriMey99},
phase conjugation \cite{GolPlaMey95,GolPlaMey96,GolMey991}, four-wave mixing
\cite{GolMey992}, etc.

In that context, the coexistence of
condensates with different magnetic quantum numbers is attractive in that it
provides a way to perform four-wave mixing experiments in collinear
geometries \cite{GolMey991}, thereby eliminating phase-matching
limitations. As an illustration, consider a Bose-Einstein condensate
of $^{23}$Na atoms in the $F=1$ hyperfine ground state, with the three
internal atomic states $|F=1,m=-1\rangle$,
$|F=1,m=0\rangle$ and $|F=1,m=1\rangle$ of degenerate energies in the
absence of external magnetic fields. The condensate is confined
by a far-off-resonant optical dipole trap. It is described by the
three-component Schr\"odinger vector field
\begin{equation}
\bbox{\Psi}({\bf r},t)
=\{\Psi_{-1}({\bf r},t),\Psi_{0}({\bf r},t),\Psi_{1}({\bf r},t)\}
\end{equation}
which satisfies the bosonic commutation relations
\begin{equation}
[\Psi_i({\bf r},t), \Psi_j^\dagger({\bf r}',t)]
=\delta_{ij}\delta({\bf r}-{\bf r}').
\end{equation}
Accounting for the possibility of two-body collisions,
its dynamics is described by the second-quantized Hamiltonian
\begin{eqnarray}
& &{\cal H}=\int d {\bf r} \bbox{\Psi}^\dagger({\bf r},t)H_0
\bbox{\Psi}({\bf r},t)
+\frac{1}{2}\int\int d {\bf r}_1 d {\bf r}_2
\nonumber\\
&\times&\bbox{\Psi}^\dagger({\bf r}_1,t)
\bbox{\Psi}^\dagger({\bf r_2},t)V({\bf r}_1-{\bf r}_2)
\bbox{\Psi}({\bf r}_2,t)\bbox{\Psi}({\bf r}_1,t),
\label{ham2}
\end{eqnarray}
where the single-particle Hamiltonian is
\begin{equation}
H_0={\bf p}^2/2M + V_{trap}
\end{equation}
and the trap potential is of the general form
\begin{equation}
V_{trap} = \sum_{m=-1}^{+1} U({\bf r})|F=1,m\rangle\langle F=1,m|.
\end{equation}
Here ${\bf p}$ is the center-of-mass momentum of the atoms of mass $M$ and
$U({\bf r})$, the  effective dipole trap potential for
atoms in the $|F=1,m \rangle$ hyperfine state, is independent of
$m$ for a non-magnetic trap.

Considering situations where the hyperfine spin $F_i = 1$ of the individual
atoms is preserved during collisions, it can be shown that in the shapeless
approximation the two-body interaction is \cite{OhmMac98,Ho98}
\begin{equation}
V({\bf r}_1-{\bf r}_2)=\hbar\delta({\bf r}_1-{\bf r}_2)
\left(c_0+c_2{\bf F}_1\cdot{\bf F}_2 \right),
\end{equation}
where $c_0 = (g_0+2g_2)/3 $ and $c_2 = (g_2-g_0)/3$, where
\begin{equation}
g_f=4\pi\hbar a_f/M,
\end{equation}
Here we label the hyperfine states of the combined system of
two collision partners with total hyperfine spin ${\bf F} = {\bf F}_1 +
{\bf F}_2$ by $|f,m \rangle$, where $f = 0, 1,2$ and $m=-f,\ldots,f$
and $a_f$ is the $s$-wave scattering length for the channel of
total hyperfine spin $f$.
Substituting this form into the second-quantized Hamiltonian (\ref{ham2})
leads to
\begin{eqnarray}
& &{\cal H}=\sum_m\int d{\bf r} \Psi_m^\dagger({\bf r},t)
\left[\frac {{\bf p}^2}{2M}+U({\bf r})\right]\Psi_m({\bf r},t)\nonumber\\
& &
+\frac{\hbar}{2}\int d{\bf r}\{(c_0+c_2)
[\Psi_1^\dagger\Psi_1^\dagger\Psi_1\Psi_1
+\Psi_{-1}^\dagger\Psi_{-1}^\dagger\Psi_{-1}\Psi_{-1}
\nonumber\\
& &
+2\Psi_0^\dagger\Psi_0
(\Psi_1^\dagger\Psi_1+\Psi_{-1}^\dagger\Psi_{-1})]
+c_0\Psi_0^\dagger\Psi_0^\dagger\Psi_0\Psi_0
\nonumber\\
& &
+2(c_0-c_2)\Psi_1^\dagger\Psi_1
\Psi_{-1}^\dagger\Psi_{-1}
\nonumber\\
& &
+2c_2(\Psi_1^\dagger\Psi_{-1}^\dagger\Psi_0\Psi_0+H.c. )\}.
\label{ham22}
\end{eqnarray}

The three terms quartic in one of the field operators
only, i.e. of the form $\Psi_i{^\dagger} \Psi_i^{\dagger} \Psi_i \Psi_i$
are self-defocussing terms, the terms involving two hyperfine states
conserve the populations of the individual spin states and merely lead to
phase shifts, and the terms involving the central mode $\Psi_0$ and
{\em both} side-modes correspond to spin-exchange collisions. This
``four-wave mixing'' interaction, involving e.g. the annihilation of a pair
of atoms with $m_F = 0$ and the creation of two atoms in the states
$m_F = \pm 1$,  leads to phase conjugation in quantum optics and to
matter-wave phase conjugation in the present case.\cite{GolMey991}

The analogy with optical four-wave mixing becomes even more apparent when
we consider a situation where atoms in the $m_F=0$ state are placed in
a linear superposition of two counterpropagating center-of-mass
modes of momenta $\pm \hbar k_0$, that is
\begin{equation}
\Psi_0(x)=\frac{1}{\sqrt V}\left(e^{ik_0 x} a_{01}+e^{-ik_0 x}a_{02}\right),
\label{cond}
\end{equation}
while the atoms of spin $m_F = \pm 1$ are taken to be the running waves
\begin{equation}
\Psi_{\pm1}(x) = \frac{1}{\sqrt V}e^{\pm ik_0x}a_{\pm 1}.
\label{side}
\end{equation}
Here $a_{01}$  and $a_{02}$ are the annihilation operators of the two
counterpropagating $m_F = 0$ modes, with $[a_{0i}, a^\dag_{0j}] =
\delta_{ij}$, $i, j = 1$ or 2, and
$a_{1},  a_{-1}$ are the corresponding operators for the running modes
associated with $m_F = \pm 1$. Finally, $V$ is the confinement
volume of the condensate. Inserting this mode expansion into the
Hamiltonian (\ref{ham22}) and ignoring all non-phase-matched contributions
yields the four-wave mixing Hamiltonian
\begin{eqnarray}
{\cal H}&=&\frac{\hbar^2k_0^2}{2M}\hat N+ \frac{\hbar c_0}{2}
{\hat N}({\hat N}-1)
\nonumber\\
&+&\frac{\hbar c_2}{2}
(a_1^\dagger a_1^\dagger a_1 a_1+a_{-1}^\dagger a_{-1}^\dagger
a_{-1} a_{-1}-2 a_1^\dagger a_1 a_{-1}^\dagger a_{-1}
\nonumber\\
&+&2 (a_1^\dagger a_1+ a_{-1}^\dagger a_{-1})
(a_{01}^\dagger a_{01}+a_{02}^\dagger a_{02})
\nonumber\\
&+&4a_1^\dagger a_{-1}^\dagger a_{01}a_{02}+4 a_1a_{-1}a_{01}^\dagger
a_{02}^\dagger),
\label{modham}
\end{eqnarray}
where we have introduced the total number of atoms
${\hat N}={\hat N}_1+{\hat N}_2$,
${\hat N}_1=a_1^\dagger a_1+a_{01}^\dagger a_{01}$ and
${\hat N}_2=a_{-1}^\dagger a_{-1} +a_{02}^\dagger a_{02}$.

This problem was solved exactly in Ref. \cite{GolMey992}
using an angular momentum
representation. Here, we reproduce just one result of this analysis
for illustration. We consider for concreteness a condensate
consisting of $N$ atoms such that there are initially
$N_1 = N_2 = N/2$, with $m \ll N/2$ atoms in the
hyperfine state $m_F = 1$ and none in the state $m_F = 0$.
The evolution of the population of the $m_F = 1$ sidemode is shown in Fig. 1
for $N=100$ atoms in the system. In case (b) the initial mode population
is $\langle a_1^\dagger a_1 \rangle = m = 5$, while case (a) illustrates
the build-up from noise, $m = 0$. In both cases, the sidemode population
exhibits an initial growth to the point where it contains about $1/3$ of the
atoms in the first case and about half of the atoms in the second. This
is followed by a collapse to a quasi-steady state population, as well as
a subsequent revival at $2c_2t_1=\pi$. This dynamics then repeats
itself periodically, with revivals at at $2c_2t_n=\pi n$, independently of
$N$.

In analogy with the optical case, matter-wave four-wave mixing is expected
to lead to the quantum entanglement of the condensate sidemodes. One can
quantify the amount of entanglement by determining the extent to which
the Cauchy-Schwartz inequality is violated by the second-order
cross-correlation functions between modes. Our results [28] show that
the correlations between central mode and sidemodes satisfy the
classical Cauchy-Schwartz inequality, while the sidemode---sidemode
cross correlations violate them. \begin{figure}
\epsfig{figure=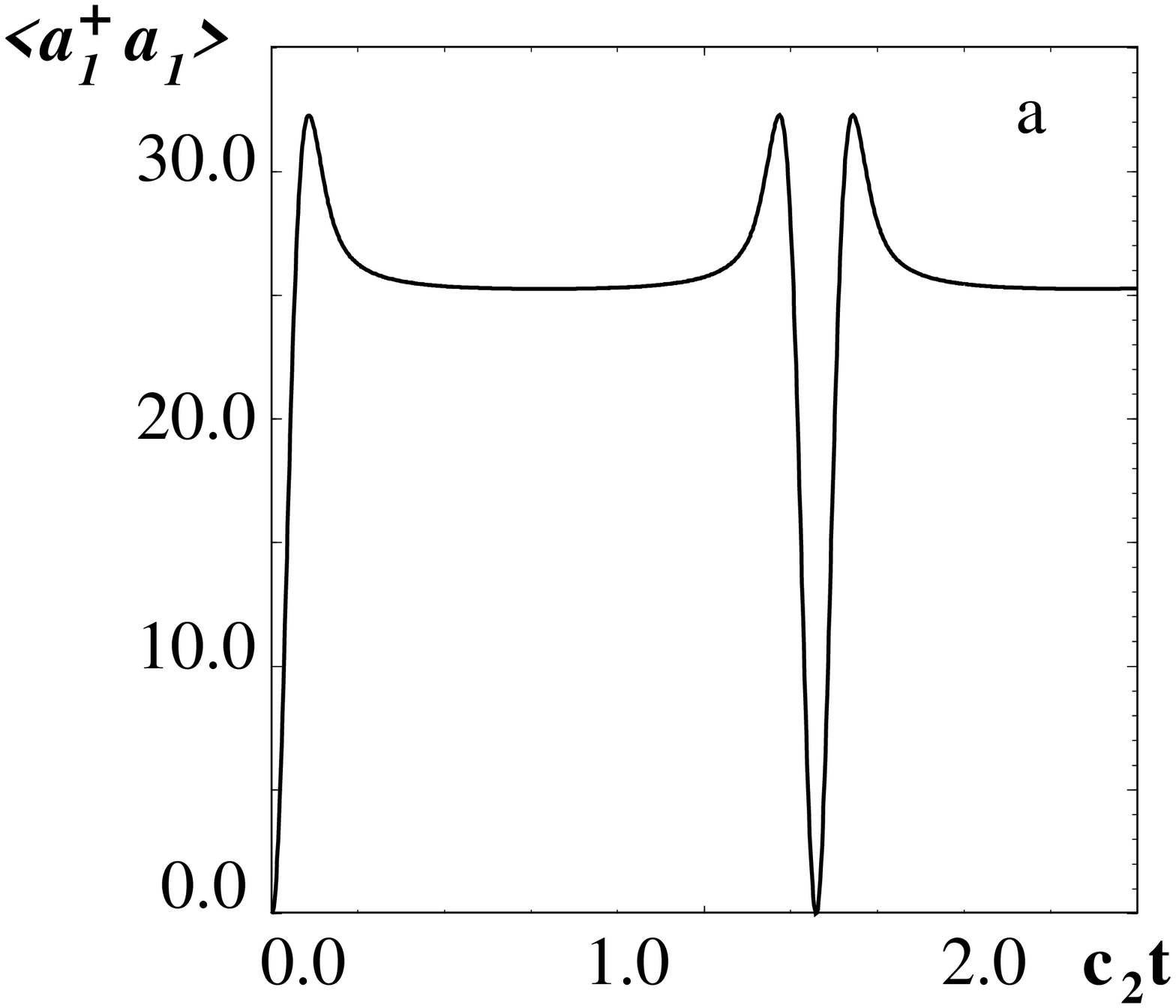,width=3.in}
\end{figure}
\begin{figure}
\epsfig{figure=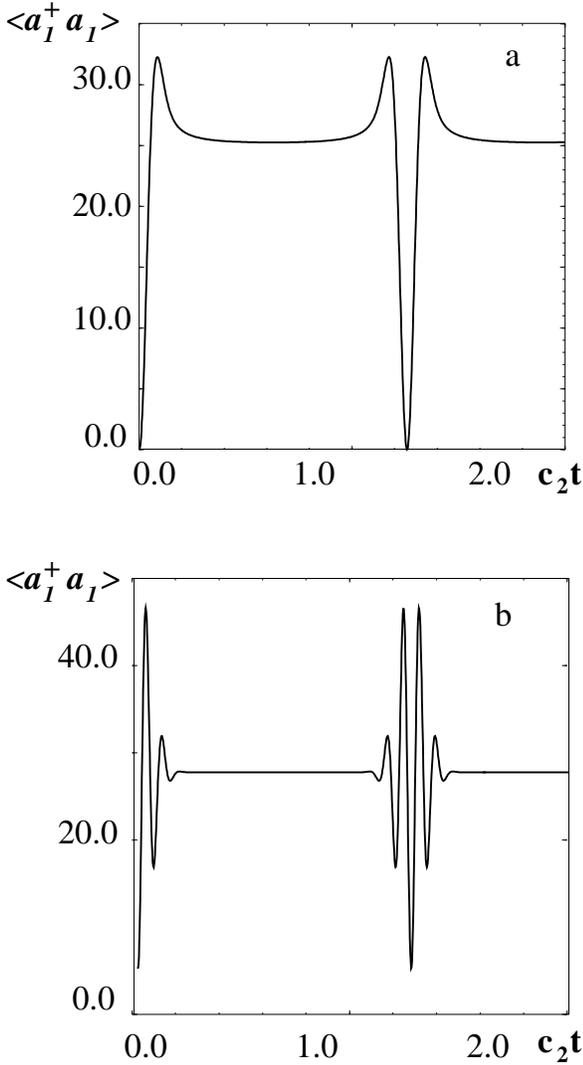,width=3.in}
\caption{
Time evolution of the side-mode population $\langle a_1^\dagger a_1\rangle$
for N=100 and initial populations (a)$\langle a_{01}^\dagger a_{01}\rangle=N/2$,
$\langle a_{02}^\dagger a_{02}\rangle=N/2$,
$\langle a_{1}^\dagger a_{1}\rangle=0$ and $\langle
a_{-1}^\dagger a_{-1}\rangle=0$ (b)
$\langle a_{01}^\dagger a_{01}\rangle=N/2-5$,
$\langle a_{02}^\dagger a_{02}\rangle=N/2$,
$\langle a_{1}^\dagger a_{1}\rangle=5$ and
$\langle a_{-1}^\dagger a_{-1}\rangle=0$.(After Ref. [28].)}
\end{figure}
The violation is particularly strong in the
case of build-up from noise, as should be intuitively expected.
The difference in the behavior of the two-mode correlation functions between
side-modes and those involving one central and one side-mode can be
intuitively understood from the form of the
wave-mixing term $a_1^\dagger a_{-1}^\dagger a_{01}a_{02}$ appearing in the
Hamiltonian (\ref{modham}). Indeed, the coupling between side-modes, involving
two annihilation operators, is reminiscent of the interaction
$a_1^\dagger a_2^\dagger$ in the Hamiltonian of parametric amplification
leading to squeezing and quantum entanglement between two sidemodes.
In contrast, the coupling between central and sidemodes involves both an
annihilation and a creation operator.

\section{Optical and matter-wave mixing}

As we have seen is section II, the matter-wave optics analog of a nonlinear optical
medium is provided by collisions. In particular, two-body collisions in the
shapeless approximation are mathematically equivalent to a local Kerr medium
with instantaneous response. These collisions result from the elimination
of the continuum of modes of the electromagnetic field from the dynamics,
very much like the elimination of the material dynamics leads to nonlinear optics.

This section turns to an intermediate regime where neither the matter waves nor
the electromagnetic field can be eliminated. In this case, it becomes in
particular possible to optically manipulate and control the quantum
statistics of matter waves (and conversely). We illustrate how this works
in the specific example of an ultracold sample of bosonic atoms located
inside an optical ring cavity and driven by a strong classical ``pump'' and
a counterpropagating weak quantized  ``probe'' optical field. Both fields
are assumed to be far off-resonant from any electronic transition, so that
all excited states can be adiabatically eliminated from the dynamics,
and the matter-wave field is effectively scalar.

The combined Hamiltonian for the atomic and probe fields is
\begin{eqnarray}
\hat{H}&=&\frac{\hbar^2}{2m}\sum_{\bf q} q^2\hat{c}^\dag({\bf q})
\hat{c}({\bf q})
+\hbar ck\hat{A}^\dag\hat{A}\nonumber\\
&+&i\frac{\hbar}{2\Delta}\sum_{\bf q}\left[g\Omega_0 e^{-i\omega_0 t}
\hat{A}^\dag\hat{c}^\dag({\bf q}-{\bf K})\hat{c}({\bf q})
- H.c.\right]\nonumber\\
&+&\frac{\hbar}{\Delta}\left(\frac{|\Omega_0|^2}{4}
+|g|^2\hat{A}^\dag\hat{A}\right)
\sum_{\bf q}{\hat c}^\dag({\bf q})\hat{c}({\bf q}).
\label{H}
\end{eqnarray}
Here, $\Omega_0$ is the Rabi frequency of the pump laser of frequency
$\omega_0$ and momentum ${\bf k_0}$, ${\hat A}$ is
the annihilation operator of the probe field of frequency
$\omega$ and momentum ${\bf k}$, satisfying $[A, A^\dag] =1$, and
${\hat c}({\bf q})$ is the annihilation operator for a ground state atom of
momentum ${\bf q}$, satisfying $[{\hat c}({\bf q}), {\hat c}^\dag({\bf q}')] =
\delta_{{\bf q},{\bf q}'}$. In addition, $\Delta$
is the detuning between the pump frequency and the upper
electronic level closest to resonance,
and $g=d[ck/(2\hbar\epsilon_0LS)]^{1/2}$ is the atom-probe coupling constant,
$d$ is the atomic dipole moment, $L$ the length of the
ring cavity, and $S$ the cross-section of the probe mode in
the region of the atomic sample. Finally, ${\bf K} \equiv {\bf k} - {\bf k_0}$
is the atomic recoil momentum resulting from the absorption of a pump photon
followed by the emission of a probe photon.

The first two terms in Eq. (\ref{H}) are the free Hamiltonians of the atomic
and probe fields, respectively. The remaining terms
correspond to the various processes by which an atom undergoes a virtual
transition under the influence of the optical fields. The first such term
involves the exchange of a photon between the pump and probe
fields, e.g. stimulated absorption of a pump photon followed by stimulated
emission of a probe photon, or vice versa. The last two terms correspond to
processes where a photon is first absorbed and then reemitted into the same field.
These transitions are recoilless, but contribute a cross-phase modulation
between the atomic and optical fields.

Assuming that the initial momentum width of the condensate is
small compared to the recoil momentum $K$, it is reasonable to treat it
as a {\em single mode} atomic field of momentum ${\bf q}=0$.
We furthermore  restrict our discussion to the case $T \ll T_c$,
where $T_c$ is the critical temperature, and assume a large condensate for
which the bare mode $q=0$ can be described to a good approximation as a
c-number, $\hat{c}(0)\to \sqrt{N}\exp(i|\Omega_0|^2t/4\Delta)$, where
$N$ is the mean number of atoms in the condensate. This
approximation neglects both the depletion which occurs as atoms are
transferred into the side-modes $q \neq 0$ and the cross-phase modulation
between the condensate and the probe field, thus it is valid for times
short enough that $\sum_{{\bf q}\neq 0}\langle\hat{c}^\dag({\bf q})
\hat{c}({\bf q})\rangle
\ll N$, and $\langle \hat{A}^\dag\hat{A}\rangle \ll |\Omega_0|^2/4|g|^2$.
This is the matter-wave optics analog of the familiar classical and undepleted
pump approximation of nonlinear optics. Hence we describe the optical and
matter-wave fields on equal footings, treating all strongly populated modes
classically and all weakly populated modes quantum-mechanically.
The growth of the system can be triggered either from vacuum
fluctuations, as we discuss in more detail shortly, or by a weak
injected probe signal.

Once we have replaced the condensate mode with its c-number counterpart, we
neglect all terms in the Hamiltonian (\ref{H}) involving the product of
three or more weakly populated modes. This yields the effective Hamiltonian
\begin{eqnarray}
\hat{H}&=&\hbar\omega_r\left[\hat{c}^\dag_+\hat{c}_+
+\hat{c}^\dag_-\hat{c}_--\delta\hat{a}^\dag\hat{a}\right.\nonumber\\
&+&\left.\chi\left(\hat{a}^\dag\hat{c}^\dag_-
+\hat{a}^\dag\hat{c}_++\hat{c}^\dag_+\hat{a}+\hat{c}_-\hat{a}\right)\right],
\label{3mH}
\end{eqnarray}
where $\omega_r=\hbar K^2/2m$, and
we have introduced the slowly varying operators
$\hat{c}_\pm=\exp(i|\Omega_0|^2t/4\Delta)\hat{c}(\pm {\bf K})$
and $\hat{a}=-i(g^\ast\Omega_0^\ast\Delta/|g||\Omega_0||\Delta|)
\exp(i\omega_0t)\hat{A}$.
The system is fully characterized by the effective coupling constant
$\chi=|g||\Omega_0|\sqrt{N}/2\omega_r\Delta$ and the dimensionless pump-probe
detuning $\delta=(\omega_0-\omega)/\omega_r$.

The Hamiltonian (\ref{3mH}) describes three coupled field modes:
the optical probe and two atomic condensate side-modes with wavenumbers
$\pm {\bf K}$.
The term $\hat{a}^\dag\hat{c}^\dag_-$ in Eq. (\ref{3mH})
describes the creation of correlated atom-photon pairs, and
immediately brings to mind
the optical parametric amplifier \cite{WalMil94}, a device
known to generate highly non-classical optical fields exhibiting
two-mode intensity correlations and squeezing, and which has been extensively
employed in the creation of entangled photon pairs for fundamental studies of
quantum mechanics, quantum cryptography and quantum computing.
A novel aspect of the present system is that it offers a way
to achieve quantum entanglement between atomic and optical fields.

We concentrate here on the equal-time two-mode intensity
cross-correlations, which are a measure of the degree of entanglement
between the modes of the system. For example, the intensity cross-correlation
function $g^{(2)}_{a-}(\tau)$ is defined as
\begin{equation}
g^{(2)}_{a-}=\frac{\langle\hat{a}^\dag(\tau)\hat{a}(\tau)
\hat{c}^\dag_-(\tau)\hat{c}_-(\tau)\rangle}
{\langle\hat{a}^\dag(\tau)\hat{a}(\tau)\rangle
\langle\hat{c}^\dag_-(\tau)\hat{c}_-(\tau)\rangle}.
\label{defg212}
\end{equation}
Other intensity cross-correlation functions such as $g^{[2]}_{a+}(\tau)$ and
$g^{(2)}_{-+}(\tau)$ are defined similarly.
For classical fields, there is an upper limit to the second-order
equal-time correlation function. It is given by the Cauchy-Schwartz inequality
\cite{WalMil94}
\begin{equation}
g^{(2)}_{ij}(\tau)\leq \left[g^{(2)}_i(\tau)\right]^{1/2}
\left[g^{(2)}_j(\tau)\right]^{1/2}.
\label{CS}
\end{equation}
Quantum mechanical fields, however, can violate this inequality
and are instead constrained by \cite{WalMil94}
\begin{equation}
g^{(2)}_{ij}(\tau)\leq\left[g^{(2)}_i(\tau)+\frac{1}{I_i(\tau)}\right]^{1/2}
\left[g^{(2)}_j(\tau)+\frac{1}{I_j(\tau)}\right]^{1/2},
\label{qmineq}
\end{equation}
which reduces to the classical result in the limit of large intensities.

Consider first the ``spontaneous'' case where the pump field is initially
in a vacuum. In this case, the equal-time intensity cross-correlation
functions are found to be
\begin{eqnarray}
g^{(2)}_{a-}&=& g^{(2)}_{-+} =
\left[2+\frac{1}{I_a(\tau)+I_+(\tau)}\right]^{1/2}
 \left[2+\frac{1}{I_-(\tau)}\right]^{1/2}, \nonumber \\
g^{(2)}_{a+} &=& 2.
\label{g212}
\end{eqnarray}
Ref. \cite{MooZobMey99} shows that both $g^{(2)}_{a-}(\tau)$ and
$g^{(2)}_{-+}(\tau)$ violate the Cauchy-Schwartz inequality, while
$g^{(2)}_{a+}(\tau)$ is consistent with classical cross-correlations.
Furthermore,  $g^{(2)}_{a-}(\tau)$ is very
close to the maximum violation of the classical inequality consistent
with quantum mechanics, whereas for $g^{(2)}_{-+}(\tau)$
the violation is not close to the allowed maximum. In the two-mode parametric
amplifier of quantum optics, the two-mode correlation function shows the
maximum violation of the Cauchy-Schwartz inequality consistent with
quantum mechanics. In the present three-mode
system, however, the two-mode cross-correlation functions involve a trace
over the third mode, hence it is not surprising that the two-mode correlations
are not maximized.

If we now allow for an injected coherent probe field, we find that in contrast
to the ``spontaneous'' case, $g^{(2)}_{a-}$ now lies somewhere in between the
quantum (\ref{qmineq}) and classical (\ref{CS}) limits. As the strength of the
classical probe field is increased, this correlation falls ever closer to the
classical upper limit, so that in the limit of very large probe fields,
one finds classical cross-correlations only.

These results show that the quantum state of momentum side-modes of a
condensate can be varied continuously
between two distinct limits by specifying the initial state of an
optical cavity mode. When it begins in the vacuum state,
the side-mode and the cavity mode fields develop with zero mean fields, thermal
intensity fluctuations, and strong quantum correlations between the modes.
In contrast, when it is prepared in strong coherent state, we approach a
``classical'' limit in which the fields develop with non-zero mean fields
having well defined phases, intensity fluctuations indicating a coherent
state, and exhibiting classical correlations only.

\section{Outlook --- Atom holography}

The interplay between optical and matter waves opens up intriguing
possibilities,  such as matter-wave holography, to which we now turn.
Optical holography can be described as the three-dimensional
reconstruction of the optical image of an arbitrarily shaped object.
Typically, this is done in a two-step process where first the
information about the object is stored in a hologram. This hologram is
created by recording, e.g., with the help of a photographic film, the
interference pattern between scattered light originating from the
illuminated
object and a (plane-wave) reference beam. The second step is the
reconstruction, which is performed by shining a reading
beam similar to the reference beam onto the hologram. The diffraction
of the reading beam from the recorded pattern yields a virtual
as well as a real optical image of the original object.

Drawing on this concept, the characteristic property of atomic
holography is that at least the final reading step is performed
with an atomic beam. In this way, an atom-optical image of the object is
created which in certain situations can be thought of as some sort
of material replica of the original. Such replicas may
have useful practical applications from atom lithography to the
manufacturing of microstructures, or quantum microfabrication.

One of the prerequisites for an actual implementation of atomic
holography is the availability of a reading beam of sufficient
monochromaticity and coherence. Given the rapid advances in atom optics
and especially in the realization of atom lasers, this requirement can be
expected to be met in the near future. One of the
greatest challenges, however, is the manufacturing of the actual hologram where
the information to be reconstructed is stored. Several schemes can be
considered. One possibility is to diffract the atoms from a mechanical mask.
The first successful realizations of such an approach have recently been
reported in Ref.\ \cite{MorYasKis96}. In these experiments the hologram was
manufactured as a binary mask written onto a thin silicon nitride
membrane. Such a hologram has the advantage of being
permanent, however, as the mask only allows for complete or
vanishing (binary) transmission of the beam at a given point one loses a
significant amount of information about the optical image. Another
interesting proposal was recently made in Ref.\ \cite{Sor97}. In
this setup the atomic beam is diffracted from the inhomogeneous light
field created by the superposition of object and reference beam. These
beams thus directly form the hologram.

In our approach, in contrast, the holographic information
is encoded into the condensate in the form of density modulations by using
writing and reference laser beams that form an optical potential for the
condensate atoms. All-atomic reading is then accomplished in a way
reminiscent of the Raman-Nath regime of diffraction between an atomic beam
and a light field.  \cite{AdaSigMly94} Specifically, the reading beam atoms,
that have a suitably chosen velocity, interact with the condensate atoms
via $s$-wave scattering and acquire a spatially dependent phase shift
reflecting the density modulations of the condensate. In the further
spatial propagation of the atoms, this phase shift gives rise to the formation
of the atom-optical image.

The idea of storing information in atomic condensates is based on the
observation that the density distribution of a condensate in the Thomas-Fermi
limit closely reflects the behavior of the confining potential. This yields
the possibility of accurate external control.
The Gross-Pitaevskii equation which governs the evolution of the
macroscopic wave function $\Phi({\bf r},t)$ describing the state of
an atomic condensate with $N$ atoms is given by \cite{DalGioPit99}
\begin{equation}
i\hbar \dot\Phi = \frac{{\bf p}^2}{2M}\Phi+V({\bf r})\Phi+
g|\Phi|^2\Phi,
\label{GP}
\end{equation}
where ${\bf p}$ denotes the atomic center-of-mass
momentum,  $M$ the atomic mass, and $V({\bf r})$ the external potential.
The strength of atomic two-body interactions is determined by
$g=4\pi\hbar^2 a/M$ with $a$ being the $s$-wave scattering length.
The normalization condition for the condensate wave function reads
\begin{equation}
\int d^3{\bf r}\,|\Phi({\bf r})|^2=N.
\label{norm}
\end{equation}
The steady state of a condensate can thus be described with a
wave function $\Phi({\bf r}, t) =\exp(-i\mu t/\hbar)\phi({\bf r})$,
with $\mu$ being the chemical potential determined by the
normalization condition Eq.\ (\ref{norm}).

In the Thomas-Fermi limit, where the effect of kinetic energy is much weaker
than the mean-field potential, the contribution of the term  ${\bf p}^2/2M$
can be neglected and the condensate density becomes
\begin{equation}
|\phi({\bf r})|^2=[\mu-V({\bf r})]/g,
\label{TFGS}
\end{equation}
From this expression we see immediately that the form of the external
potential is replicated in the density profile of the atomic condensate.

For reading, we consider an all-atomic
scheme which has the fundamental advantage of allowing one to reconstruct
a {\em material} ``replica'' of the stored object. Specifically, the
reading beam is a monoenergetic atomic beam of velocity ${\bf v}_{rd}$
impinging at some angle onto the condensate. We assume that the
internal state of these incoming atoms is such that they are only weakly
perturbed by the writing and trap potentials, so that their dominant
interaction is scattering by the atoms in the condensate. It is important at
this point to emphasize that the atoms in the reading beam {\em need not}
be of the same species as the condensate atoms. In principle they
could be of just about any element or even molecule.

Fig.\ 2 shows an example, taken from Ref. \cite{ZobGolMey99},
which demonstrates
explicitly that an original rectangular aperture can be indeed reconstructed
with our proposed atom holography method.

\begin{figure}
\epsfig{figure=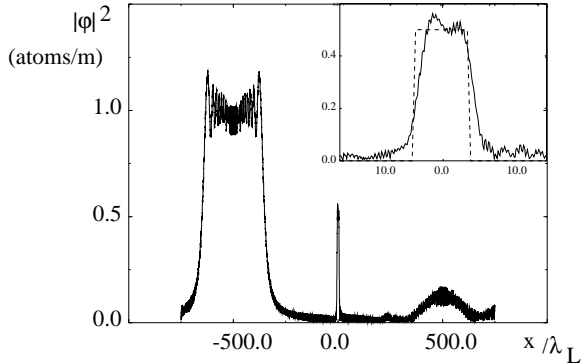,width=3.in}
\caption{Reconstructing a replica of the original object from the atomic
hologram. The reading beam consists of a monochromatic beam of sodium atoms
moving at an angle $\beta_A$ from the $z$ axis at a velocity
$v_{rd}=0.1$ m/sec. Shown is the atomic density profile at a distance
$\Delta z$ from the condensate such that the quadratic phase shift of the
conjugate image is precisely canceled. The insert compares the reconstructed
and original objects. The off-axis feature for positive $x$ corresponds to
the real object, for which the quadratic phase is still present. The
large off-axis feature at negative $x$ is the background. (After Ref. [35].)}
\end{figure}

\noindent\acknowledgements
This work is supported in part by the U.S.\ Office of Naval Research
under Contract No.\ 14-91-J1205, by the National Science Foundation under
Grant No.\ PHY98-01099, by the U.S.\ Army Research Office, and by the
Joint Services Optics Program.

\end{document}